\documentstyle[11pt,newpasp,twoside]{article}
\def\edcomment#1{\iffalse\marginpar{\raggedright\sl#1\/}\else\relax\fi}
\reversemarginpar

\begin{document}
\title{Two more years of GRB optical afterglows}
\vspace{-.5cm}
\author{Nicola Masetti}
\affil{IASF/CNR, via Gobetti 101, I-40129 Bologna, Italy}

\vspace{-.4cm}
\begin{abstract}
I review the main observational achievements
concerning GRB optical afterglows occurred in the past two years
Their emission seems to be well interpreted in terms of the `fireball'
model, although a variety of behaviours are observed.
Evidence for supernova-like late time `bumps' in the optical light curves
is growing; however, other interpretations are possible.
Fast follow-up observations allowed the detection of faint optical
afterglows, and seem to suggest that a fraction of the so-called `dark'
bursts might actually be re-classified as `dim' GRBs.
Perspectives for future observations in this field are also outlined.
\end{abstract}

\vspace{-1cm}
\section{Introduction: what happened in the past two years}

\vspace{-.3cm}
I will summarize here the discoveries 
accomplished during the past two years in the optical domain
and concerning the Gamma-Ray Burst (GRB) afterglow phenomenon,
along with the information carried by these detections.

I start by briefly listing, in a (not too tight) chronological order, what
in my opinion are the technical and instrumental landmarks in the GRB
field since October 2000. First, by the end of 2000, {\it HETE-2} became
fully operational and started delivering fast GRB triggers; meanwhile, the
GRB community began making wider use of robotic telescopes
(among them ROTSE, LOTIS, BOOTES, BART, KAIT, Palomar 48-inch, Mount
Stromlo 50-inch, RAPTOR).
In parallel, afterglow searches were performed also by
applying new techniques such as the color-color diagram (Rhoads 2001) 
or the image subtraction (Alard 2000) methods. On April 2002 the
{\it Mars Odyssey} spacecraft joined the {\it InterPlanetary Network}
({\it IPN}), making this satellite network active and efficient again for
GRB localizations after the end of the {\it NEAR} mission on February
2001. The bad news of this two-year period is the end of {\it
BeppoSAX} operative life on April 30, 2002, exactly six years after its
launch. A happy appendix is however the successful launch, on October 17,
2002, of {\it INTEGRAL}, which will be able to provide real-time
arcmin-sized GRB localizations.

\vspace{-.6cm}
\section{The new discoveries}

\begin{figure}
\plotfiddle{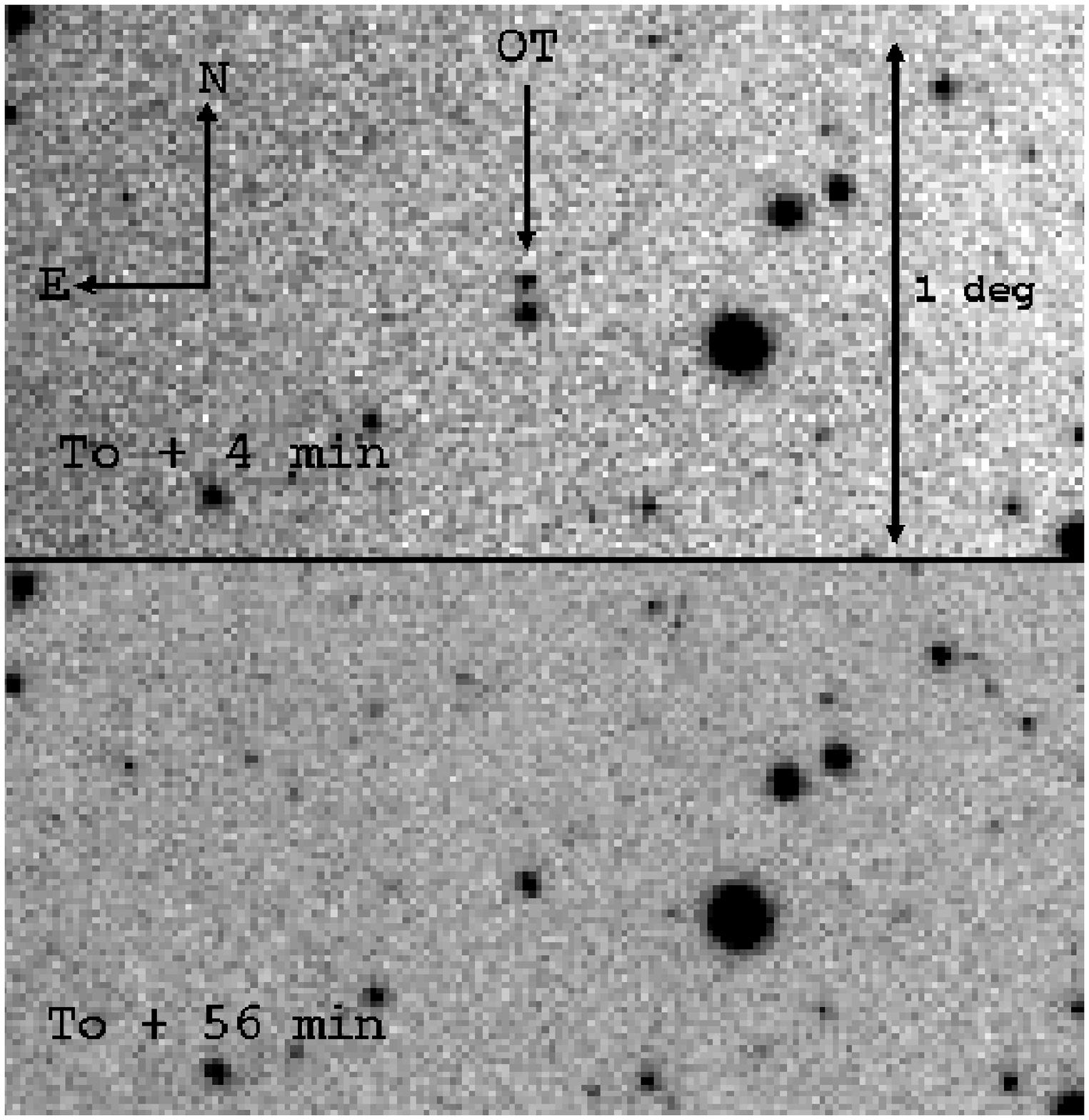}{4cm}{0}{30}{30}{-190}{-80}
\plotfiddle{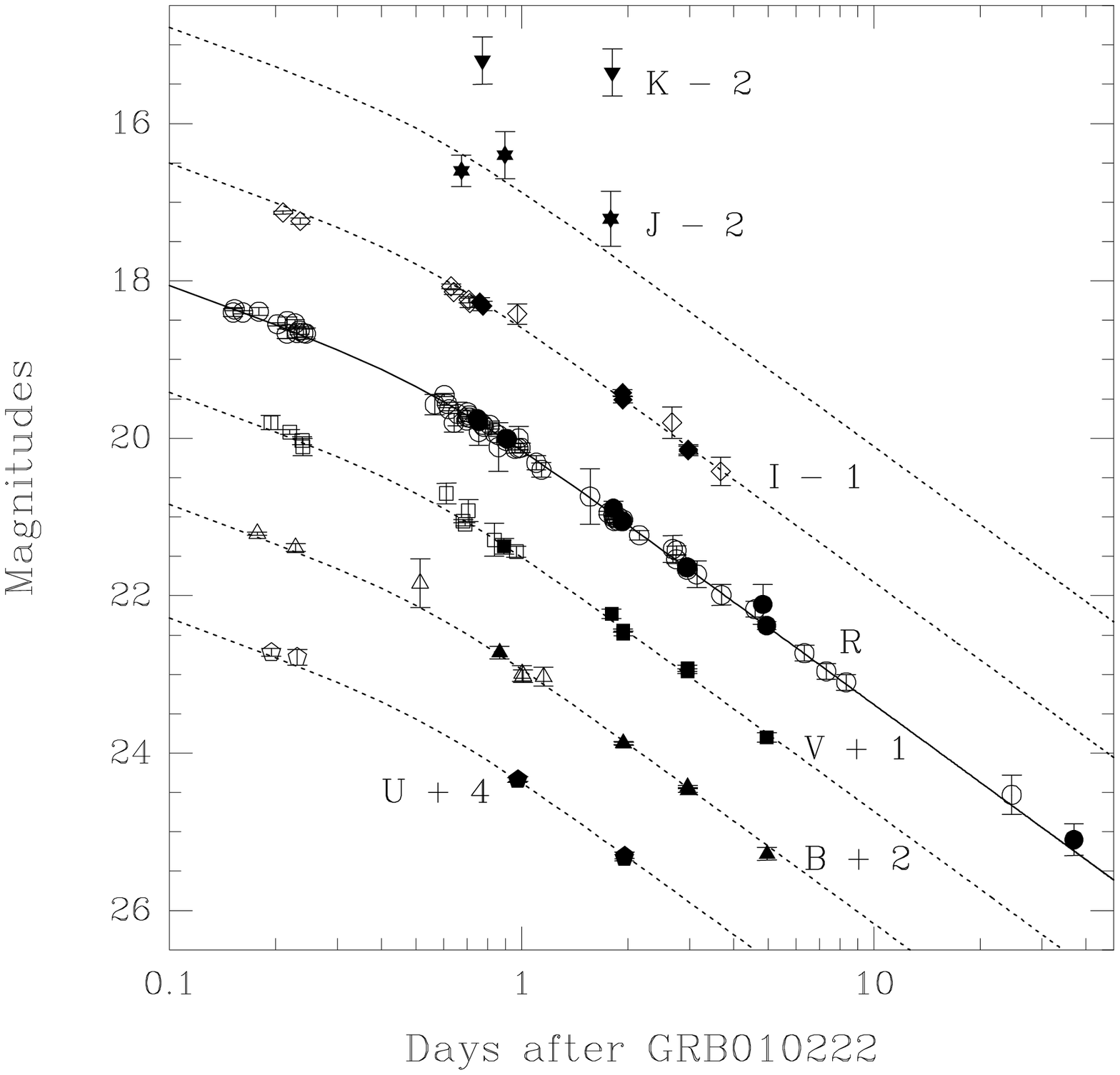}{4cm}{0}{35}{35}{-10}{20}
\vspace{-3.2cm}
\caption{{\it Left}: BOOTES images showing the fading of the possible 
GRB000313 OT (adapted from Castro-Tirado et al. 2002).
{\it Right}: Optical and near-infrared light curves of the
GRB010222 afterglow (from Masetti et al. 2001).}
\end{figure}

\vspace{-.3cm}
In this section I will review 
the main characteristics of the Optical Transients (OTs) discovered in the 
period October 2000 - September 2002, along with a GRB, exploded in March 2000, 
which can potentially be very interesting and for which a comprehensive study
was published only very recently. I will also briefly mention the new OT 
discoveries occurred after the end of this Workshop and before the
proceedings submission deadline (January 2003).
In the following I will denote with $\alpha$ and $\beta$ the time decay
and spectral indices of the afterglow emission, respectively, according to
the common notation $F(t,\nu) \propto t^{-\alpha} \nu^{-\beta}$.

\smallskip

{\sl GRB000313}. This short/hard GRB was detected by {\it BATSE} only,
thus it had not an accurate localization. Nevertheless, within the
3$\sigma$ BATSE error box, an object (Fig. 1, left) of magnitude $I$ =
9.4$\pm$0.1 was detected by BOOTES about 4 minutes after the GRB. 
Subsequent observations showed that neither an optical afterglow, nor a
host down to a limiting magnitude $R \sim$ 24.5 were detected at the
position of the transient. If this was indeed the optical counterpart of
this GRB, the early light curve had a decay index $\alpha >$ 2.2, similar
to the decay observed in the optical flash of GRB990123 (Akerlof et al.
1999). Thus, BOOTES may have caught the first optical flash associated
with a short/hard GRB. The absence of an afterglow and of a host would
suggest that indeed short/hard GRBs occur in low-density environments
outside their hosts. A detailed study on this GRB can be found in
Castro-Tirado et al. (2002).

\smallskip

{\sl GRB001007}. This GRB was detected by the {\it IPN} and its OT was
characterized by a very fast ($\alpha \sim$ 2) light curve decay.
Its non-detection in LOTIS data suggested a possible early break in the
light curve. The spectral and temporal behaviour of the OT were plausibly
modeled assuming a collimated emission.
The host galaxy was detected at $R \sim$ 24.8 and presented an irregular
shape. More information is available in Castro-Cer\'on et al. (2002).

\smallskip

{\sl GRB001011}. The OT of this GRB localized by {\it BeppoSAX} was the
first to be found with the color-color diagram technique (Rhoads 2001)
extended to the infrared data, and later confirmed by its fading.
It showed a time decay with $\alpha \sim$ 1.3 and was located in a host
with $R \sim$ 25.4. Further details on the discovery of this OT can be
found in Gorosabel et al. (2002a).

\smallskip

{\sl GRB010222}. This GRB, one of the best shots by {\it BeppoSAX},
was accompanied by a relatively bright OT with three absorption
systems in the spectrum at redshifts 0.927, 1.155 and 1.475 (Jha et al.
2001; Masetti et al. 2001). Its light curve underwent an achromatic
break (with the exception of the $K$ band; Fig. 1, right) $\sim$0.7 days
after the GRB, while the broadband spectrum suggested that the cooling
frequency $\nu_{\rm c}$ was in the UV domain $\sim$1 day after the prompt
event (Masetti et al. 2001). No significant polarization was detected
(Bj\"ornsson et al. 2002). A possible further light curve break was
found by Fruchter et al. (2001a) $\sim$10 days after the GRB.
No unanimous interpretation of the fireball shape, physics and emission
mechanisms at play in this OT was reached: subrelativistic transition
(Masetti et al. 2001), continuous energy injection (Bj\"ornsson et al.
2002), low electron distribution index $p$ (Sagar et al. 2001) were
invoked to explain the OT behaviour.
High-resolution spectra indicated that the GRB environment was not
particularly dense (Mirabal et al. 2002); this result is at variance with
the findings of Masetti et al. (2001) and Savaglio et al. (2003).

\smallskip

{\sl GRB010921}. This was the first GRB detected by {\it HETE-2} for which
an OT was found, and the first OT found by means of image subtraction 
technique (Price et al. 2001); it would have probably been overlooked 
if this method was not applied. It was at redshift $z$ = 0.451
(Price et al. 2001) and its light curve showed two breaks: one within
1 hour (Park et al. 2001) and the other about 35 days (Price et
al. 2002a) after the GRB. Indication for spherical fireball expansion and
for substantial absorption in the host were found (Price et al. 2001). 
Any underlying supernova (SN), if present, was at least 1.3 mag fainter
than SN1998bw (Price et al. 2002a). 

\smallskip

{\sl GRB011211}. The OT of this {\it BeppoSAX} GRB was located at $z$ =
2.14 (Fruchter et al. 2001b). The light curve showed a break $\sim$2 days
after the GRB and fluctuations which could be due to circumburst medium
(CBM) inhomogeneities on a scale of tens of AUs; observations are
consistent with the OT emission being produced by a fireball expanding in
a homogeneous medium (Holland et al. 2002). The OT was not polarized
(Covino et al. 2002a).

\smallskip

\begin{figure}
\plotfiddle{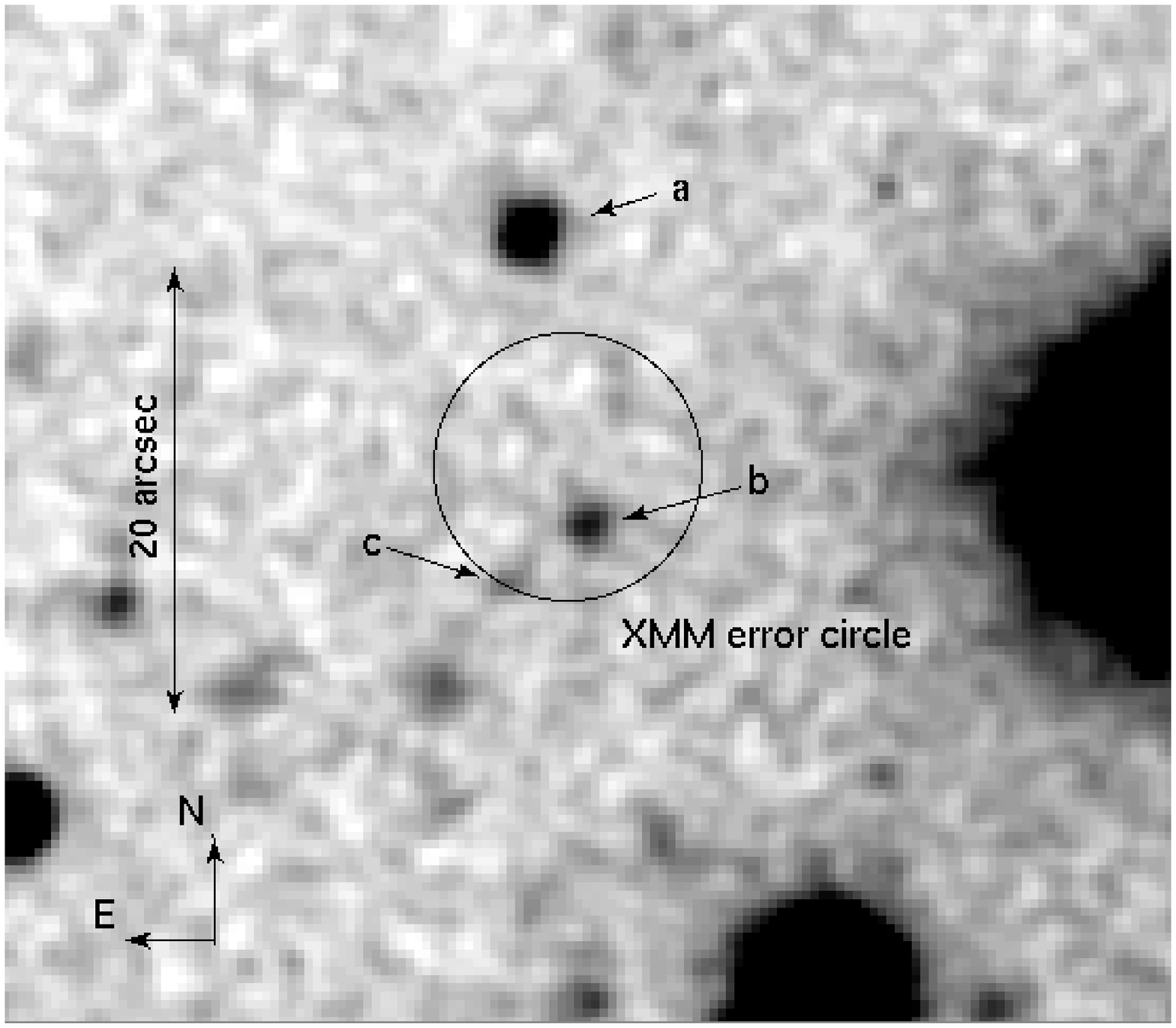}{3.5cm}{0}{32}{32}{-180}{-100}
\plotfiddle{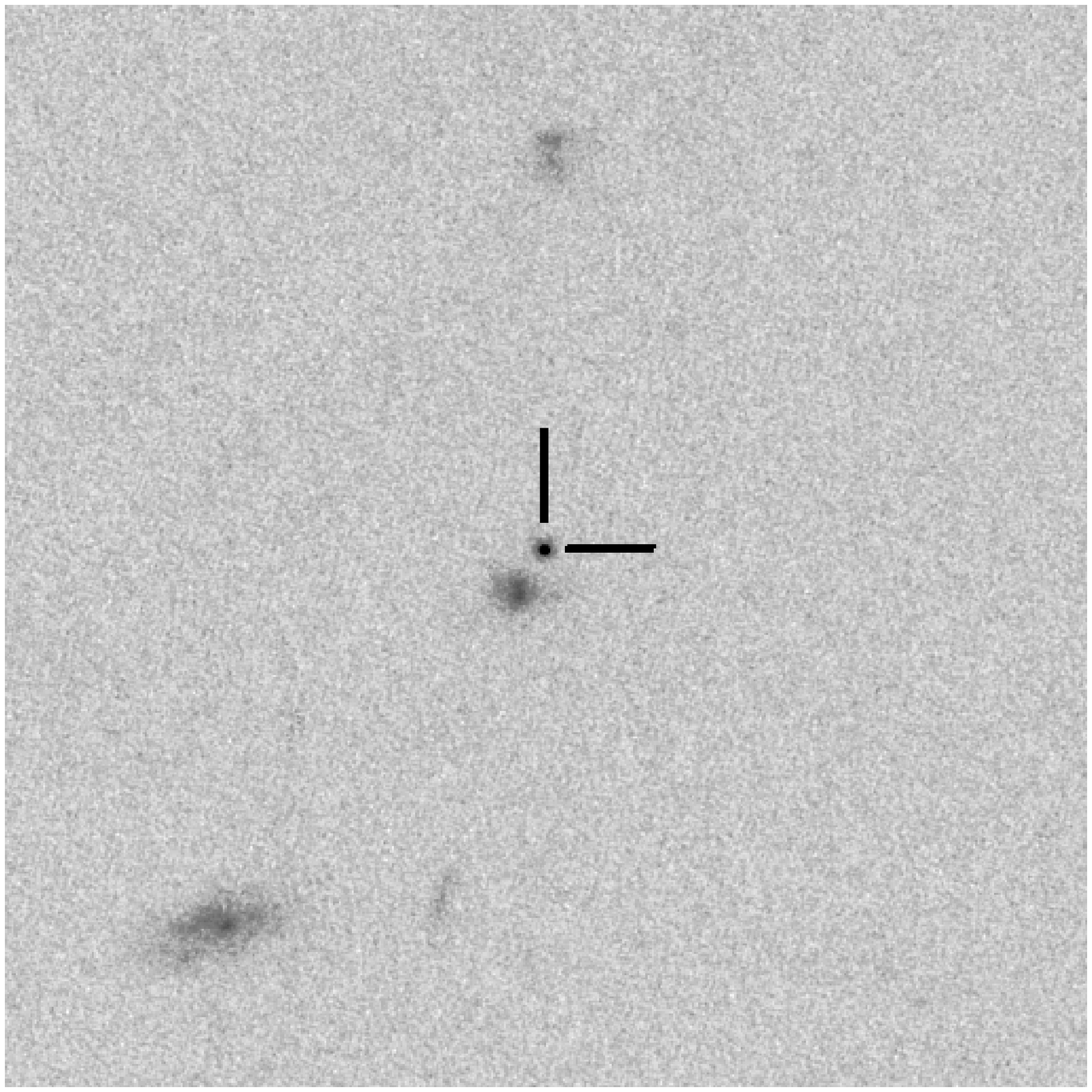}{3.5cm}{0}{30}{30}{10}{20}
\vspace{-2.8cm}
\caption{{\it Left}: {\it XMM-Newton} error box of GRB020322. The OT is
the object marked with the letter `b' (from Bloom et al. 2002a).
{\it Right}: HST image of the GRB020410 OT. The object is indicated
by the tick marks. Adapted from Fruchter et al. (2002c).}
\end{figure}

{\sl GRB020305}. This was the second OT detected following an 
{\it HETE-2} alert. 
It showed decay and spectral indices $\alpha \sim$ 1.3 and $\beta \sim$
1, respectively (Lee et al. 2002), and the presence of a host was reported
(Ohyama et al. 2002; Gorosabel et al. 2002b).
Further observations can be found in Yoshida et al. (these proceedings).

\smallskip

{\sl GRB020322}. The OT of this {\it BeppoSAX} GRB was one of the dimmest
ones detected up to now: 8 hours after the GRB it was already at
magnitude $R \sim$ 23.3 (Bloom et al. 2002a; Fig. 2, left) and was
very likely discovered only thanks to the arcsecond-sized position
available through the {\it XMM-Newton} detection of the X--ray afterglow
(Ehle et al. 2002). A possible break around $\sim$1 day after the GRB was
detected in its optical light curve (Burud et al. 2002). 

\smallskip

{\sl GRB020331}. The {\it HETE-2} `Easter GRB' was associated with a very
slowly decaying OT ($\alpha \sim$ 0.6; Fox et al. 2002). This was possibly
due to the contamination of the underlying host (Dullighan et al. 2002).
However, Castro-Cer\'on et al. (these proceedings) reported a flattening
and a possible subsequent restart of the (slow) decay occurring $\sim$1
week after the GRB, which would suggest that this behaviour might be
intrinsic and not due to the host contribution.

\smallskip

{\sl GRB020405}. The OT of this {\it IPN} GRB, at a redshift $z$ = 0.69
(Masetti et al. 2002; Price et al. 2002b), showed a bump in the light
curve between 10 and 20 days after the GRB: this was interpreted by Price
et al. (2002b) as due to an underlying SN. Besides, the OT was possibly
highly polarized ($P \sim$ 10\%; Bersier et al., 2002; however, see
the review by Covino in these proceedings). A fairly complete
spectrophotometric and polarimetric study of this GRB, along with an
alternative interpretation of the optical light curve bump, can be found
in Masetti et al. (these proceedings).

\smallskip

{\sl GRB020410}. One of the last {\it BeppoSAX} localizations allowed the
detection of an OT, for the first time ever, by means of HST (Fruchter
et al. 2002c; Fig. 2, right). This was possible thanks to the very
accurate (20$''$ in radius) X--ray afterglow position by {\it BeppoSAX} so
that the entire error box could be fitted in the HST-WFPC2
field of view. Comparison with ground-based observations showed that the
OT light curve underwent a break; however, the possibility that this was
actually a high-$z$ SN cannot be excluded (Fruchter et al. 2002c). 

\smallskip

{\sl GRB020813}. This was one of the brightest GRB detected by {\it
HETE-2} up to now. Its OT showed two absorption systems in its spectrum,
at redshifts 1.122 and 1.254 (Price et al. 2002c). A marked light curve
break was detected $\sim$1 day after the GRB (Bloom et al. 2002b), while a
flattening, most likely due to the presence of the host, was observed at
late times (Gorosabel et al. 2002). For the first time in the case of an
OT, a very high resolution spectrum was acquired (Fiore et al. 2002).
Optical polarization, possibly variable, was also detected (Barth et al.
2003; Covino et al. 2002b; Rol et al., these proceedings). 

\smallskip

{\sl Other OTs}. After the end of the meeting, and before the submission
deadline (January 15, 2003), the OTs of three more GRBs (021004, 021211
and 030115, all localized by {\it HETE-2}) were detected. The fast
localizations of 021004 and 021211 allowed the detection, by means of
robotic telescopes, of the corresponding OTs within the first minutes
after the GRB trigger: the emission was in both cases interpreted as the
tail of the optical flash simultaneous with the GRB itself (Kobayashi \&
Zhang 2003; Wei 2003). The OT of GRB021004 was quite bright and well
sampled; observations suggest a clumpy CBM (e.g. Lazzati et al.
2002). GRB021211 was instead associated with a quite faint OT (Fox et al.
2003) and, possibly, an even fainter OT was found for GRB030115 (Masetti
et al. 2003) at the position of the infrared afterglow (Levan et al.
2003). 

A controversial case is that of the {\it HETE-2} X--ray flash XRF020903. 
In its error box a decaying optical source showing a SN-like bump after
$\sim$30 days was detected at $z$ = 0.25 (Soderberg et al. 2002). Thus,
were this indeed the OT of XRF020903, it would be the first optical
counterpart ever detected for an XRF; moreover, its behaviour would not
substantially differ from that of GRB OTs, in spite of their different
high-energy prompt event characteristics. Alternatively, an AGN
interpretation for this OT was given (Gal-Yam 2002); recent HST
observation showed however that the variable object is in a complex of
interacting galaxies, but not at the center of any of them (Levan et al.
2002). 

\smallskip

I intentionally did not mention the OTs of GRB011121 and GRB020124: they
will be discussed in these proceedings by Greiner, Zeh et al. and Price.

\vspace{-.6cm}
\section{Summary: the lessons we (hopefully) learnt}

\vspace{-.3cm}
The discoveries of new OTs during the past two years increased our 
knowledge (and in some cases our uncertainties) on the behaviour and
emission mechanisms of GRB afterglows in the optical.
We found that OTs show no ``preference" in their expansion 
geometry, fireball evolution or environmental characterstics.

Thanks to the availability of accurate (arcmin-sized) real-time GRB 
alert and to the increasing number of fast-response robotic telescopes,
we are getting closer and closer to see the OT behaviour in its very
first phases. Recent observations are moreover suggestive
of a noteworthy point: there indeed seems to be wide a spread in the OT
brightnesses and this might, at least partially, explain the `dark'
optical afterglow conundrum. That is, quite likely some of them are
actually intrinsically `dim' OTs, rather than obscured or high-$z$
ones.

Also, evidence for late-time bumps in the OT light curves is growing.
These are customarily associated with the emerging contribution of a SN
exploded at, or around, the time of the GRB; however, this behaviour is
not universal (see e.g. the case of GRB010921), and other interpretations
such as dust echoes (Esin \& Blandford 2000), late shell collisions (Kumar
\& Piran 2000) or ion-proton shell interactions (Beloborodov 2003) should
be explored. 

What is the lesson we should learn from all this? In my opinion, it can be
listed in six points (some of them, admittedly, a bit tautologic): 
1) {\it we need fast high-energy arcsec localization} to pinpoint the GRB 
position and start very deep serches as soon as the alert is distributed.
This can also help to quickly find `dim' and/or rapidly decaying OTs.
The {\it Swift} satellite will be able to fulfill this task; 2) as
{\it HETE-2} and {\it INTEGRAL} can already provide real-time GRB alerts,
{\it we need very fast-response robotic telescopes}. The {\it REM}
telescope (Zerbi et al., these proceedings), starting on March 2003, will 
secure observational coverage beginning within 10 seconds after the GRB
trigger; 3) sometimes, error boxes with sizes of several arcmin of
radius are provided. In this case, {\it the use of telescopes with
large-field imagers should be encouraged}; 4) as some afterglows can be
optically faint due to local dust obscuration or high-$z$ effects, 
{\it wider use of near-infrared imaging should be made}. This also will
be addressed by {\it REM}; 5) {\it the use of new techniques for OT
searches} (such as color-color diagram, image subtraction, along with the
development of pipelines for fast reduction  of wide-field images) {\it
should also be encouraged}; 6) finally, and when applicable, {\it
co-operation among groups} (and possibly from GRBs also...) would be
desirable to expand our understanding of these still enigmatic
and therefore fascinating phenomena.

\end{document}